\documentclass{ws-procs961x669}            
\def\be{\begin{equation}}
\def\ee{\end{equation}}
\def\bq{\begin{eqnarray}}
\def\eq{\end{eqnarray}}
\def\beq{\begin{eqnarray}}
\def\eeq{\end{eqnarray}}

\def\a{\alpha}
\def\b{\beta}

\def\pa{\partial}
\begin{document}
\title{Brane-world asymptotics in a nonlinear fluid bulk}

\author{I. Antoniadis}

\address{Laboratoire de Physique Th\'{e}orique et Hautes Energies - LPTHE,
Sorbonne Universit\'{e}, CNRS 4 Place Jussieu, 75005 Paris, France, and\\
Institute for Theoretical Physics, KU Leuven,\\ Celestijnenlaan 200D, B-3001 Leuven, Belgium\\
E-mail: antoniad@lpthe.jussieu.fr}

\author{S. Cotsakis}

\address{Institute of Gravitation and Cosmology, RUDN University\\
ul. Miklukho-Maklaya 6, Moscow 117198, Russia, and\\
Research Laboratory of Geometry,  Dynamical Systems  and Cosmology,\\
University of the Aegean, Karlovassi 83200, Samos, Greece\\
E-mail: skot@aegean.gr}

\author{Ifigeneia Klaoudatou}

\address{Research Laboratory of Geometry,  Dynamical Systems  and Cosmology,\\
University of the Aegean, Karlovassi 83200, Samos, Greece \\
E-mail: iklaoud@aegean.gr}

\begin{abstract}
We present recent results on the asymptotics of a brane-world that
consists of a flat 3-brane embedded in a five-dimensional bulk. The bulk matter is modelled by a fluid
that satisfies a nonlinear equation of state of the form
$p = \gamma\rho^{\lambda}$, where p is the ‘pressure’ and $\rho$ is the
‘density’ of the fluid. We show that for appropriate ranges of the parameters $\gamma$
and $\lambda$, it is possible to construct a regular solution, compatible with
energy conditions, that successfully localizes gravity on the brane.
These results improve significantly previous findings of the study of a bulk fluid with
a linear equation of state.
\end{abstract}


\bodymatter
\section{Introduction}
In this paper, we review very recent work, presented in \cite{ack7}, on the
asymptotic behaviors of a class of brane-worlds. The models consist of a
flat 3-brane embedded in a five-dimensional bulk filled with a fluid that
satisfies a nonlinear equation of state $p = \gamma\rho^{\lambda}$,
where p is the ‘pressure’ and $\rho$ is the ‘density’ of the fluid and $\gamma$,
$\lambda$ are constants.

Such an equation of state, has been studied in cosmology for its contribution
in avoiding big-rip singularities during late time asymptotics
\cite{srivastava, diaz, nojiri}, in obtaining inflationary models with special
properties \cite{ba90}, in unifying models of dark energy and dark matter
\cite{ka, sen}, as well as in studies of singularities \cite{ck1, ck2, not}.

Our goal in studying these brane-world models, is to find a solution that is
regular and meets physical requirements set by energy conditions and
localization of gravity on the brane. In our previous work in
\cite{ack1, ack2, ack3, ack4, ack5, ack6}, which extended work done in
brane-worlds with scalar fields \cite{nima, silverstein, gubser}, we studied a
$3$-brane (flat or curved) and a fluid with a linear equation of state $p=\gamma\rho$.
We found that such a solution becomes possible only for the special value of $\gamma=-1$ and
only for a flat brane.
These results gave us the motivation to explore whether the
generalization of the linear equation of state to a nonlinear one,
could have a decisive impact in finding a solution with all required properties
and valid for a wide range of $\gamma$ and $\lambda$.
This exploration turned out to be very fruitful, and we were able to find in \cite{ack7}
such solutions for a flat brane and a fluid with $\gamma<0$ and $\lambda>1$.

As the analysis of the nonlinear equation of state is quite complicated, for the most
part of this paper, we will focus on the study of a specific value of $\lambda=3/2$,
which is simpler for illustration and offers important insight on the behavior of
solutions for general $\lambda>1$.

The structure of this paper is the following:
In Section 2, we set up our model and derive the field equations.
In Section 3, we formulate the null energy condition which we
embody in our solutions later in Section 4. In Section 5, we derive a
matching solution for the specific value of $\lambda=3/2$ and based
on this solution, we calculate, in Section 6, the four-dimensional Planck mass. Finally, in
Section 7, we discuss briefly the asymptotic behaviors of solutions for general
$\lambda$.
\section{Setup of brane-world}
To set up our model, we start by considering a flat 3-brane embedded in a five-dimensional
bulk with a metric of the form,
\be
\label{warpmetric}
g_{5}=a^{2}(Y)g_{4}+dY^{2},
\ee
where $a(Y)$ is the warp factor 
and $g_{4}$ is the four-dimensional flat metric,
{\it i.e.},
\be
\label{branemetrics}
g_{4}=-dt^{2}+dx_1^2+dx_2^2+dx_3^2.
\ee
The timelike coordinate is denoted by $t$ and the spacelike
ones by $(x_1,x_2,x_3,Y)$. In our notation, Capital Latin indices are taken as
$A,B,\dots=1,2,3,4,5$ while lowercase Greek indices range as $\a,\b,\dots =1,2,3,4$. The 5-dimensional
Riemann tensor is defined by the formula,
\be
R^{A}_{\,\,\,BCD}=\pa_{C}\Gamma^{A}_{\,\,\,BD}-\pa_{D}\Gamma^{A}_{\,\,\,BC}+\Gamma^{M}_{BD}\Gamma^{A}_{MC}-\Gamma^{M}_{BC}\Gamma^{A}_{MD}
\ee
the Ricci tensor is the contraction,
\be
R_{AB}=R^{C}_{\,\,\,ACB},
\ee
and the five-dimensional Einstein equations on the bulk space are given by,
\be
G_{AB}=R_{AB}-\frac{1}{2}g_{AB}R=\kappa^{2}_{5}T_{AB}.
\ee

Next, we set up the bulk matter component by considering a bulk fluid with
an energy-momentum tensor of the form,
\be
\label{T old}
T_{AB}=(\rho+p)u_{A}u_{B}-p g_{AB},
\ee
where $p$ is the `pressure' and $\rho$ is the `density' which
we take as functions only of the extra dimension $Y$.
In (\ref{T old}), the velocity vector field is $u_{A}=(0,0,0,0,1)$, that is
$u_{A}=\pa/\pa Y$, parallel to the $Y$-dimension. We assume that the pressure and density are
interconnected by the following nonlinear equation of state
\be
\label{eos}
p=\gamma \rho^{\lambda},
\ee
with $\gamma$ and $\lambda$ constants.

The Einstein equations can be written as
\bq
\label{syst2i}
\frac{a'^{2}}{a^{2}}&=&\frac{\kappa_{5}^{2}}{6}\rho,\\
\label{syst2ii}
\frac{a''}{a}&=&-\frac{\kappa_{5}^{2}}{6}{(2\gamma\rho^{\lambda}+\rho)},
\eq
with the prime $(')$ denoting differentiation with respect to $Y$.
The equation of energy-momentum conservation, on the other hand,
$$\nabla_{A}T^{AB}=0,$$
becomes
\be
\label{syst2iii}
\rho'+4(\gamma\rho^{\lambda}+\rho)\frac{a'}{a}=0.
\ee

At this point, it is useful to clarify the terminology that we use in what follows.
We use the term {\em finite-distance singularities} to refer to the following singular
behaviors of $a$, occurring within a finite distance $Y_{s}$:
\begin{itemize}
\item $a\rightarrow 0^{+}$, as $Y\rightarrow Y_{s}$ (collapse singularity),

\item $a\rightarrow\infty$, as $Y\rightarrow Y_{s}$ (big-rip singularity).
\end{itemize}
With $Y_{s}$, we denote a finite value of $Y$ designating the position of the
singularity. The above behaviors may be accompanied, in general,
by a divergence in the density, or, even in the pressure of the fluid.
Note that, these singularities are not related to geodesic incompleteness
as in standard cosmology, but rather on a pathological behavior of the warp factor.
In the absence of finite-distance singularities, we call the solutions regular
and include in this category the behaviors of the warp factor given above, provided
that these occur {\em only} at infinite distance, {\em i.e} $Y\rightarrow\pm \infty$.

We are interested to see, if it is possible to find solutions of
Eqs.~(\ref{syst2i})-(\ref{syst2iii}) that rectify previous findings of solutions of
brane-worlds with linear bulk fluids studied in \cite{ack1, ack2, ack3, ack4, ack5, ack6}.
As mentioned in the introduction, excepting the case of a flat brane and a fluid with
$\gamma=-1$, the main problem
that we faced in the study of a linear bulk fluid, is that regular solutions
that exist in the case of curved branes, or even singular ones out of which we
can construct a regular matching solution for both
flat and curved branes, lead to a challenging compromise: they either satisfy
energy conditions, or, localize gravity on the brane. Therefore, a solution
arising from the consideration of this different and more complicated type of
fluid would be preferable and worth examining in detail, if it has each one of the following characteristics:
\begin{itemize}
\item it is regular (no finite-distance singularities)
\item it satisfies physical conditions, such as energy conditions
\item it leads to a finite four-dimensional Planck mass, thus, it localizes gravity on the brane.
\end{itemize}

We can simplify the situation of finding such a solution, by incorporating, for
example, the energy conditions a priori. As we show in the next
Sections this is indeed possible. The regular nature as well as the requirement
of a finite four-dimensional Planck mass can be checked as a subsequent step, since to accomplish that,
we first need to find the exact form of the warp factor and density.
\section{The null energy condition}
We are going to derive the null energy condition for our type of nonlinear
bulk fluid described by (\ref{T old}) and (\ref{eos}) and transform it to a condition imposed on the density
$\rho$ and parameters $\gamma$ and $\lambda$ of the equation of state.

To achieve this, we follow the technique analysed in \cite{ack4, ack6}. We start by recalling that the metric (\ref{warpmetric}) and
the bulk fluid (\ref{T old}) and (\ref{eos}) depend only on the extra dimension $Y$ and
they are therefore static with respect to the time coordinate $t$. Such
fluid can be viewed as an anisotropic one with energy-momentum tensor
\be
\label{T new}
T_{AB}= (\rho^{0}+p^{0})u_{A}^{0}u_{B}^{0}-p^{0}g_{\alpha\beta}\delta_{A}^{\alpha}\delta_{B}^{\beta}-
p_{Y}g_{55}\delta_{A}^{5}\delta_{B}^{5},
\ee
where $u_{A}^{0}=(a(Y),0,0,0,0)$, $A,B=1,2,3,4,5$ and $\alpha,\beta=1,2,3,4$.
As a next step, we compare (\ref{T old}) with (\ref{T new}) and find that the following
relations should hold
\bq
\label{p y to rho}
p_{Y}&=&-\rho\\
\label{rho new}
\rho^{0}&=&-p\\
\label{p new}
p^{0}&=&p.
\eq
We note that once we combine the last two relations we get
\be
\label{p new to rho new}
p^{0}=-\rho^{0},
\ee
which means that this type of matter satisfies a cosmological
constant-like equation of state.
Substituting (\ref{p y to rho})-(\ref{p new to rho new})
in (\ref{T new}), we find that
\be
T_{AB}= -p g_{\alpha\beta}\delta_{A}^{\alpha}\delta_{B}^{\beta}+
\rho g_{55}\delta_{A}^{5}\delta_{B}^{5}.
\ee

Here, we focus only on the null energy condition according to which, every
future-directed null vector $k^{A}$ should satisfy \cite{poisson}
\be
T_{AB}k^{A}k^{B}\geq 0.
\ee
This means that we should have
\be
p+\rho\geq 0,
\ee
from which we find
\be
\label{nec-0}
\gamma\rho^{\lambda}+\rho\geq 0,
\ee
after inputting the equation of state $p=\gamma\rho^{\lambda}$.
Now, Eq.~(\ref{nec-0}) can be written as
\be
\rho^{\lambda}(\gamma+\rho^{1-\lambda})\geq 0
\ee
and since $\rho\geq 0$ from Eq.~(\ref{syst2i}), we arrive at the
the final form of the null energy condition, which reads
\be
\label{nec}
\gamma+\rho^{1-\lambda}\geq 0.
\ee
In the next Section, we are going to incorporate this condition in
the process of deriving our solutions.
\section{Regular solution}
We start solving the system of Eqs.~(\ref{syst2i})-(\ref{syst2iii}),
by first integrating the continuity equation (\ref{syst2iii}) to find the
relation between the warp factor and the density. In the integration process
we arrive at a logarithmic term of the form
$\ln |\gamma+\rho^{1-\lambda}|$. To incorporate from the beginning the null
energy condition (\ref{nec}), we choose to ignore the absolute value and simply
put this term equal to $\ln (\gamma+\rho^{1-\lambda})$. The resulting relation
between $\rho$ and $a$ is
\be
\label{rho to a}
\rho={(-\gamma+c_{1}a^{4(\lambda-1)})}^{1/(1-\lambda)},
\ee
where
\be c_1=\frac{\gamma+\rho_0^{1-\lambda}}{a_0^{4(\lambda-1)}},
\ee
with $\rho_0=\rho(Y_0), a_0=a(Y_0)$ being the initial conditions.
According to (\ref{nec}) this translates to $c_{1}\geq 0$.

It is crucial to note that there is a singularity in the density
($\rho\rightarrow\infty$)
for $\lambda>1$ and
\be
\label{sing}
a^{4(\lambda-1)}=\dfrac{\gamma}{c_{1}}.
\ee
However, it is possible to avoid this singularity, by restricting the
range of the parameter $\gamma$, only to negative values, which we do in the
analysis that follows.

Substituting Eq.~(\ref{rho to a}) in Eq.~(\ref{syst2i}) and integrating, we obtain,
\be
\label{hyper_integral}
\int\dfrac{a}{(c_{1}-\gamma a^{4(1-\lambda)})^{1/(2(1-\lambda))}}\,da=\pm\dfrac{\kappa_{5}}{\sqrt{6}}\int dY.
\ee
The complicated integral on the LHS of Eq. (\ref{hyper_integral}), can be
greatly simplified for those values of $\lambda$ that make $1/(2(1-\lambda))$
a negative integer. This is possible when $\lambda=(n+1)/n$, with $n=2k$ and
$k$ a positive integer.

Clearly, the simplest case is for $n=2$ which corresponds to $\lambda=3/2$.
Then $1/(2(1-\lambda))=-1$ and we can directly integrate Eq. (\ref{hyper_integral})
to get the solution in the following implicit form
\be
\label{sol_3/2}
\pm Y+c_{2}=\dfrac{\sqrt{6}}{\kappa_{5}}\left(\dfrac{c_{1}}{2} a^{2}-\gamma\ln a\right),
\ee
where $c_{2}$ is an integration constant. It is straightforward to see that
this solution is regular: all possible singular behaviors of $a$, namely
$a\rightarrow 0$ and $a\rightarrow \infty$ happen only for infinite
values of $Y$ with $p$ and $\rho$ either vanishing there, or, approaching
finite nonzero values. In particular, we have
\beq
\label{a_div_3/2}
a&\rightarrow&\infty, \quad\rho\rightarrow  0,\quad p\rightarrow  0, \quad \textrm{as}\quad Y\rightarrow\pm\infty\\
\label{a_0_3/2}
a&\rightarrow&0^{+}, \quad  \rho\rightarrow 1/\gamma^{2},\quad p\rightarrow -1/\gamma^{2},\quad \textrm{as}\quad Y\rightarrow\pm\infty.
\eeq

Summarizing, the solution for $\lambda=3/2$ has already two out of the three
good qualities mentioned in the previous Section: it is regular and at the same time
satisfies the null energy condition. In the following section, we construct an
appropriate matching solution out of this solution, which leads to a finite
four-dimensional Planck mass, thus fitting perfectly the profile of a plausible solution.

As for the asymptotics for general values of $\lambda$, it is
interesting to note here that even though the case of $\lambda=3/2$ is the
simplest possible one for the nonlinear fluid, further analysis performed in
\cite{ack7}, shows that this case is significant in the sense that it can act
as a guide for the whole range of values of $\lambda>1$. This follows from
the fact that all of its good physical properties are, in a way, inherited
to all cases with $\lambda>1$.
\section{Matching solution for $\lambda=3/2$}
For $\lambda=3/2$, a matching solution that leads to a finite four-dimensional
Planck mass,
can be constructed in the following way:
First, we write down the two different branches of solution of Eq.~(\ref{sol_3/2}),
\beq
\label{pos_branch}
Y^{+}&=&h_{+}(a)=\dfrac{\sqrt{6}}{\kappa_{5}}\left(\dfrac{c_{1}^{+}}{2}a^{2}-\gamma\ln a\right)+C_{2}^{+}\\
\label{neg_branch}
Y^{-}&=&h_{-}(a)=\dfrac{\sqrt{6}}{\kappa_{5}}\left(-\dfrac{c_{1}^{-}}{2}a^{2}+\gamma\ln a\right)+C_{2}^{-},
\eeq
using the notation $Y^{\pm}$ to describe the solutions
for the $(\pm)$ choice of sign in Eq.~(\ref{sol_3/2}). Similarly, the notation
$c_{1}^{\pm}$, $C_{2}^{\pm}$ describes the values of $c_{1}$ and $\mp c_{2}$,
respectively, on the $(\pm)$ branch of $Y$.

Second, in order to gain a complete idea of the behavior of $h_{+}(a)$, we
study the shape of the curve of $h_{+}(a)$ by calculating the first and second derivative of
$h_{+}(a)$. These read
\beq
\label{h+'}
h_{+}'(a)&=&\dfrac{\sqrt{6}}{\kappa_{5}}\left(\dfrac{c_{1}^{+}a^{2}-\gamma}{a}\right)\\
\label{h+''}
h_{+}''(a)&=&\dfrac{\sqrt{6}}{\kappa_{5}}\dfrac{c_{1}^{+}}{a^{2}}\left(a^{2}+\dfrac{\gamma}{c_{1}^{+}}\right).
\eeq
For our choice of integration constants and parameters, $c_{1}^{+}>0$ and
$\gamma<0$, Eq.~(\ref{h+'}) shows that $h_{+}$ is strictly increasing
for all values of $a$. Also, Eq.~(\ref{h+''})
shows that the graph of $h_{+}$ changes from concave downward on
$(0,\sqrt{-\gamma/c_{1}^{+}})$ to concave upward on $(\sqrt{-\gamma/c_{1}^{+}},\infty)$, making the point
$(\sqrt{-\gamma/c_{1}^{+}},-\sqrt{6}/\kappa_{5}(\gamma/2+\gamma \ln \sqrt{-\gamma/c_{1}^{+}})+C_{2}^{+})$,
an inflection point of the graph of $h_{+}$.

On the other hand, the first two derivatives of $h_{-}(a)$ are exactly the
opposites of the ones of $h_{+}$ given in Eq.~(\ref{h+'}) and Eq.~(\ref{h+''}) above. This means that $h_{-}$ is strictly decreasing and
the graph of $h_{-}$ changes from concave upward on $(0,\sqrt{-\gamma/c_{1}^{-}})$ to concave downward on $(\sqrt{-\gamma/c_{1}^{-}},\infty)$,
with $(\sqrt{-\gamma/c_{1}^{-}},\sqrt{6}/\kappa_{5}(\gamma/2+\gamma \ln \sqrt{-\gamma/c_{1}^{-}})+C_{2}^{-})$
being an inflection point of the graph of $h_{-}$.

We can match the two branches at their common inflection point by making an
appropriate choice of the constants $c_{1}^{\pm}$ and $C_{2}^{\pm}$. This is a third
step in the matching process. Naturally, we assume that the warp factor is
continuous
\be
\label{match cond1}
\sqrt{\dfrac{-\gamma}{c_{1}^{+}}}=\sqrt{\dfrac{-\gamma}{c_{1}^{-}}},\quad \textrm{or}
,\quad c_{1}^{+}=c_{1}^{-}=c_{1}
\ee
and that, of course, the inflection point has the same $Y$ coordinate
through the two branches, which leads to
\be
h_{+}\left(\sqrt{\dfrac{-\gamma}{c_{1}}}\right)=h_{-}\left(\sqrt{\dfrac{-\gamma}{c_{1}}}\right)
\ee
and yields the following relation between $C_{2}^{+}$ and $C_{2}^{-}$,
\be
\label{match cond2}
-\dfrac{\sqrt{6}}{\kappa_{5}}\left(\dfrac{\gamma}{2}+\gamma \ln \sqrt{\dfrac{-\gamma}{c_{1}}}\right)+C_{2}^{+}=\dfrac{\sqrt{6}}{\kappa_{5}}\left(\dfrac{\gamma}{2}+\gamma \ln \sqrt{\dfrac{-\gamma}{c_{1}}}\right)+C_{2}^{-},
\ee
or,
\be
C_{2}^{+}=\dfrac{\sqrt{6}}{\kappa_{5}}\left(\gamma+2\gamma\ln \sqrt{\dfrac{-\gamma}{c_{1}}}\right)+C_{2}^{-}.
\ee
There is an axis of symmetry at $Y=Y_{s}$, with $Y_{s}$ given by
\be
\label{Y_s}
Y_{s}=h_{\pm}(\sqrt{-\gamma/c_{1}})=-\dfrac{\sqrt{6}}{\kappa_{5}}\left(\dfrac{\gamma}{2}+\gamma \ln \sqrt{\dfrac{-\gamma}{c_{1}}}\right)+C_{2}^{+}=\dfrac{\sqrt{6}}{\kappa_{5}}\left(\dfrac{\gamma}{2}+\gamma \ln \sqrt{\dfrac{-\gamma}{c_{1}}}\right)+C_{2}^{-}.
\ee
Since the embedding of the brane introduces a $Y\rightarrow-Y$ symmetry in the bulk,
it is natural to position the brane at $Y=Y_{s}$. Without loss of generality and in order
to simplify our expressions, we can place the brane at $Y=0$, that is
we take $Y_{s}=0$ in $(\ref{Y_s})$. This will set the values of
$C_{2}^{\pm}$ to
\be
\label{C_2_3/2}
C_{2}^{+}=\dfrac{\sqrt{6}}{\kappa_{5}}\left(\dfrac{\gamma}{2}+\gamma \ln \sqrt{\dfrac{-\gamma}{c_{1}}}\right)=-C_{2}^{-}.
\ee
The matching solution described by Eq. (\ref{pos_branch}) and (\ref{neg_branch}) that satisfies the above boundary condition
can be written as
\be
\label{matching sol 3/2}
|Y|=\dfrac{\sqrt{6}}{\kappa_{5}}\left(-\dfrac{c_{1}}{2}a^{2}+\gamma\ln a-\dfrac{\gamma}{2}-\gamma \ln \sqrt{\dfrac{-\gamma}{c_{1}}}\right),\quad 0<a\leq\sqrt{\dfrac{-\gamma}{c_{1}}}.
\ee
It is essential that the density is well defined and continuous at the position of the brane
$Y=0$. Here, we find
\be
\label{rho match 3/2}
\rho\left(\sqrt{\dfrac{-\gamma}{c_{1}}}\right)\equiv\rho(0)=\dfrac{1}{4\gamma^{2}},
\ee
where $\rho(0)$ denotes $\rho(Y=0)$ and it is an abbreviation of
$\rho(h_{\pm}^{-1}(\sqrt{-\gamma/c_{1}}))$.

\begin{figure}[h]
\begin{center}
\includegraphics[scale=1.0]{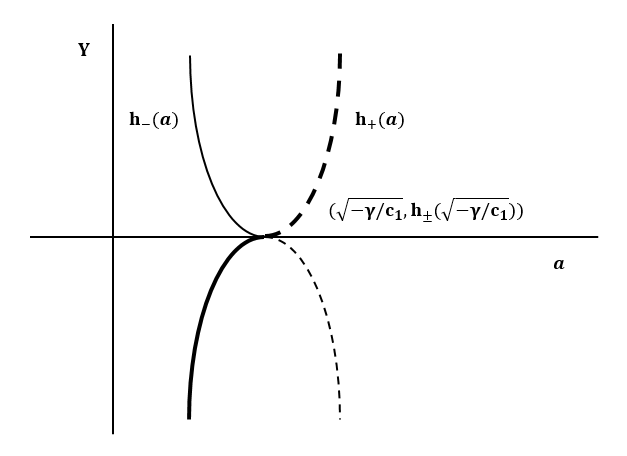}
\caption{Matching graphs of $h_{\pm}(a)$ for $\lambda=3/2$.}
\label{matching graphs}
\end{center}
\end{figure}

\begin{figure}[h]
\begin{center}
\includegraphics[scale=1.0]{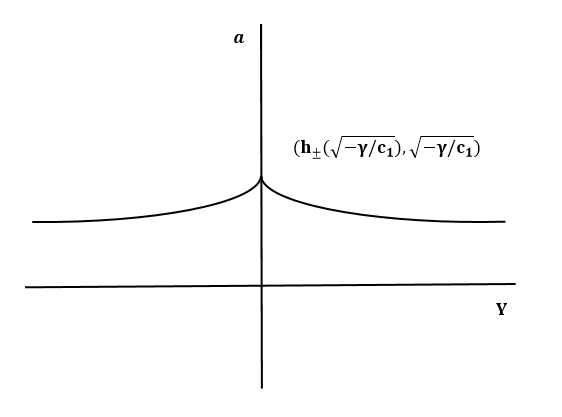}
\caption{Graph of $a(Y)$ for the matching solution
for $\lambda=3/2$.}
\label{matching graphs rotated}
\end{center}
\end{figure}

The graph of $h_{+}$ is depicted in Fig.~\ref{matching graphs} with a thick solid and
dashed line, while the graph of $h_{-}$, in the same Figure, is depicted with a
thin solid and dashed line. The matching solution is build up only from the
solid parts of both graphs, as this is the only choice that leads to a solution
with a finite four-dimensional Planck mass. In Fig.~\ref{matching graphs rotated}, we have rotated the axes of
Fig.~\ref{matching graphs}, for a more convenient view of the evolution
of the warp factor as a function of $Y$.

Finally, we take into account the jump of the derivative of
the warp factor across the brane and find, for our type of geometry, the following
junction condition
\be
\label{match_3}
a'(0^{+})-a'(0^{-})=2a'(0^{+}) =-\kappa_{5}^{2}\dfrac{f(\rho(0))a(0)}{3},
\ee
where $f(\rho(0))$ is the tension of the brane. Using
Eq.~(\ref{matching sol 3/2}) in the above equation we can deduce
the form of the brane tension, this reads
\be
\label{match_4}
f(\rho(0))=-\dfrac{\sqrt{6}}{2\kappa_{5}\gamma}=\dfrac{\sqrt{6\rho(0)}}{\kappa_{5}},
\ee
where the last equality is derived by using Eq.~(\ref{rho match 3/2}).

This completes the construction of a viable matching solution for
$\lambda=3/2$. In the next Section, we are going to show that this solution also leads
to a finite four-dimensional Planck mass.
\section{Localization of gravity for $\lambda=3/2$}
To calculate the value of the four-dimensional Planck mass,
$M_{p}^{2}=8\pi/\kappa$, we use the following integral (see \cite{ack4, ack6, ack7, forste}),
\be
\label{mass}
\frac{\kappa_{5}^{2}}{\kappa}=\int_{-Y_{c}}^{Y_{c}}a^{2}(Y)dY.
\ee
We can deduce the behaviour of $a^{2}$ as $Y\rightarrow-\infty$ from
Eq.~(\ref{matching sol 3/2}). We find
\be
\label{sq a sol 3/2}
a^{2}\sim e^{-(\sqrt{6}\kappa_{5}/(3\gamma))Y},
\ee
where the symbol $\sim$ is used to signify `behaves according to', and
using the symmetry of Eq.~(\ref{matching sol 3/2}), we can write the above integral in the following form
\be
\int_{-Y_{c}}^{Y_{c}}a^{2}dY=2\int_{-Y_{c}}^{0}a^{2}dY\sim
2\int_{-Y_{c}}^{0}e^{-(\sqrt{6}\kappa_{5}/(3\gamma))Y}dY=
-\sqrt{6}\dfrac{\gamma}{\kappa_{5}} (1-e^{(\sqrt{6}\kappa_{5}/(3\gamma))Y_{c}}).
\ee
Taking $Y_{c}\rightarrow\infty$ and keeping in mind that $\gamma<0$, it follows
that the four-dimensional Planck mass remains finite and is proportional to
$$-\sqrt{6}\dfrac{\gamma}{\kappa_{5}}.$$

In conclusion, we have seen a specific example of the nonlinear equation
of state $p=\gamma\rho^{3/2}$, for which it is possible to construct a regular
matching solution compatible with the null condition and which also localizes
gravity on the brane. In the discussion below, we outline what happens for
extended ranges of values of $\lambda$.
\section{Discussion}
We briefly review in this Section, the behavior of solutions for general
$\lambda$, a detailed analysis can be found in \cite{ack7}.

The key factor that determines whether a solution of
Eqs.~(\ref{syst2i})-(\ref{syst2iii}), is regular, or, singular,
is the ordering of $\lambda$ with respect to the value one:
for $\lambda<1$, we encounter singular solutions, whereas, for
$\lambda>1$ the solutions are regular.

Before presenting the solutions for general $\lambda$, we emphasize that
for $\lambda\neq 1+1/(2k)$, where $k$ is a positive integer, the integral on
the LHS of Eq.~(\ref{hyper_integral}) cannot in general be calculated
directly. However, we can still deduce the behaviors of solutions by expressing
them in terms of the Gaussian hypergeometric function $_{2}F_{1}(\alpha,b,c;z)$.

In particular, for $\lambda<1$, the solution is given implicitly in terms of
a hypergeometric function and reads
\be
\label{a_sing}
\pm Y+c_{2}=\frac{\sqrt{6}}{2\kappa_{5}}c_{1}^{1/(2(\lambda-1))}a^{2}\,_{2}F_{1}\left(\dfrac{1}{2(1-\lambda)},\dfrac{1}{2(1-\lambda)},
\dfrac{1}{2(1-\lambda)}+1;\dfrac{\gamma}{c_{1}}a^{4(1-\lambda)}\right).
\ee
As shown in \cite{ack7}, this solution has a finite-distance singularity at
$Y\rightarrow \pm c_{2}$, with
\beq
& & a\rightarrow 0^{+}, \quad \rho\rightarrow\infty, \quad p\rightarrow 0, \quad \textrm{if}\quad
\lambda<0\\
& & a\rightarrow 0^{+}, \quad \rho\rightarrow\infty, \quad p\rightarrow \infty,\quad \textrm{if}\quad
0<\lambda<1.
\eeq
The behaviors of $p$ and $\rho$ above, have been deduced from Eq.~(\ref{eos})
and Eq.~(\ref{rho to a}).

For $\lambda>1$, on the other hand, we have the following types of solution.
First, we have the general form of solution for $\lambda=1+1/(2k)$,
with $k$ a positive integer, which is
\be
\label{sol_endpoints}
\pm Y+c_{2}=\dfrac{\sqrt{6}}{\kappa_{5}}\left(\sum_{s=0}^{k-1}\dfrac{k!}{(k-s)!s!}\dfrac{c_{1}^{k-s}}{2-2s/k} a^{2-2s/k}(-\gamma)^{s}+(-\gamma)^{k}\ln a\right).
\ee
Second, for $\lambda>3/2$ we have the following solution
\beq
\label{sol_3/2 above}
\nonumber
\pm Y+c_{2}&=&\dfrac{\sqrt{6}}{\kappa_{5}}\left(\dfrac{a^{2}}{2}(c_{1}-\gamma a^{4(1-\lambda)})^{1/(2(\lambda-1))}-
 \dfrac{\gamma c_{1}^{(3-2\lambda)/(2(\lambda-1))}}
{2(3-2\lambda)}a^{2(3-2\lambda)}\times \right. \\
& &\left.
_{2}F_{1}\left(\dfrac{3-2\lambda}{2(1-\lambda)},\dfrac{3-2\lambda}{2(1-\lambda)},\dfrac{3-2\lambda}{2(1-\lambda)}+1;\dfrac{\gamma}{c_{1}}a^{4(1-\lambda)}\right)\right),
\eeq
Finally, the case of $1<\lambda<3/2$ is more complicated: we have forms of
solutions valid for $1+1/(2k)<\lambda<1+1/2(k-1)$, with $k$ a positive integer
such that $k\geq2$, and given by
\beq
\label{sol_n+1/n_n-1/n-2}
\nonumber
\pm Y+c_{2}&=&\frac{\sqrt{6}}{\kappa_{5}}\left(\sum_{s=0}^{n/2-1}\dfrac{(-\gamma)^{s}}{2(1-2s(\lambda-1))}\left(c_{1}a^{4(\lambda-1)}-\gamma\right)^{1/(2(\lambda-1))-s}\right.\\ \nonumber
&+&\left.\dfrac{(-\gamma)^{n/2}(c_{1})^{((n+1)-n\lambda)/(2(\lambda-1))}}{2((n+1)-n\lambda)}a^{2((n+1)-n\lambda)}\times \right.\\
&\times&\left._{2}F_{1}\left(\dfrac{(n+1)-n\lambda}{2(1-\lambda)},\dfrac{(n+1)-n\lambda}{2(1-\lambda)},\dfrac{(n+1)-n\lambda}{2(1-\lambda)}+1;\dfrac{\gamma}{c_{1}}a^{4(1-\lambda)}\right)\right).
\eeq

All solutions for $\lambda>1$ are free from finite-distance singularities, and
their asymptotic behaviors are
\beq
\label{a_div}
a&\rightarrow&\infty, \quad \rho\rightarrow  0,\quad p\rightarrow  0, \quad \textrm{as}\quad Y\rightarrow\pm\infty\\
\label{a_0}
a&\rightarrow&0^{+}, \quad \rho\rightarrow(-\gamma)^{1/(1-\lambda)},\quad p\rightarrow -(-\gamma)^{1/(1-\lambda)},\quad \textrm{as}\quad Y\rightarrow\pm\infty.
\eeq
The procedure for deriving a matching solution with a finite four-dimensional Planck
mass, presented in the previous Sections, can still be performed for
all solutions with $\lambda>1$ and can be found in \cite{ack7}.

Summarizing, the effect of the nonlinear equation of state is catalytic
in the process of creating a regular solution with essential physical properties
and valid for a wide range of parameters.
It is left to be seen from a physics point of view, what are the possible field realizations that could lead to such
an equation of state.
\section*{Acknowledgments}
Work partially performed by I.A. as International professor of the Francqui Foundation, Belgium.
\bibliographystyle{ws-procs961x669}
\bibliography{ws-pro-sample}

\end{document}